\documentclass[twocolumn]{jpsj311}
\usepackage{txfonts}
\usepackage{url}

\title{New Nonequilibrium-to-Equilibrium Dynamical Scaling and \\
Stretched-Exponential Critical Relaxation in Cluster Algorithms}

\author{Yoshihiko Nonomura}
\inst{Computational Materials Science Unit, 
National Institute for Materials Science, Tsukuba, Ibaraki 305-0044, Japan} 

\abst{
Nonequilibrium relaxation behaviors in the Ising model on 
a square lattice based on the Wolff algorithm are totally 
different from those based on local-update algorithms. 
In particular, the critical relaxation is described by the 
stretched-exponential decay. We propose a 
novel scaling procedure to connect nonequilibrium 
and equilibrium behaviors continuously, and find that 
the stretched-exponential scaling region in the Wolff 
algorithm is as wide as the power-law scaling region 
in local-update algorithms. We also find that relaxation 
to the spontaneous magnetization in the ordered phase 
is characterized  by the exponential decay, not the 
stretched-exponential decay based on local-update 
algorithms.
}

\begin{document}
\maketitle

{\it Introduction.} 
In Monte Carlo study of phase transitions, the critical slowing 
down has been a serious problem. As long as local-update 
algorithms are used, equilibration at the critical temperature 
$T_{\rm c}$ is practically impossible for fairly large systems 
because of a large dynamical critical exponent $z$. In order 
to overcome this difficulty, two kinds of approaches have been 
proposed. One is the cluster algorithms,\cite{SW,Wolff1} 
in which global update based on a percolation theory is 
introduced and the exponent $z$ is considered to be greatly 
reduced. The other is the nonequilibrium relaxation 
(NER) method,\cite{NERrev} in which the critical relaxation 
in local-update algorithms is analyzed with the dynamical 
scaling theory\cite{MSdyn} and equilibration is avoided.

The initial aim of the present study was to integrate the 
two approaches and to develop a more efficient scheme. 
However, when early-time relaxation in cluster 
algorithms at $T_{\rm c}$ is precisely investigated, 
we find that the standard NER method based on 
the dynamical scaling theory no longer holds. 
That is, early-time relaxation is not described by the 
power-law decay but by the stretched-exponential 
decay in cluster algorithms. The NER method has been 
used to estimate the exponent $z$, and the results of 
a study of Ising models starting from the all-up state 
based on cluster algorithms\cite{Du} can be compared 
with the present results, even though a size dependence 
of the relaxation time $\tau$ of the energy was observed 
in the former. The previously-revealed logarithmic 
size dependence of $\tau$, which suggests $z=0$, 
might be consistent with our finding, although 
a stretched-exponential time dependence of 
physical quantities was not explicitly shown previously.

The outline of the present Letter is as follows. 
After a brief overview of the formulation of numerical analyses,
we first exhibit the stretched-exponential decay at $T_{\rm c}$. 
In order to confirm this nontrivial behavior, we propose a 
new nonequilibrium-to-equilibrium scaling analysis and 
compare this behavior with the established power-law 
decay using the Metropolis algorithm at $T_{\rm c}$. 
We also investigate the relaxation behavior 
using the Wolff algorithm below $T_{\rm c}$. 
The background of the stretched-exponential decay and 
the relation with previous studies of the dynamical critical 
exponent $z$ are discussed, and future tasks of the 
present framework are proposed.

\begin{figure}[b]
\includegraphics[width=88mm]{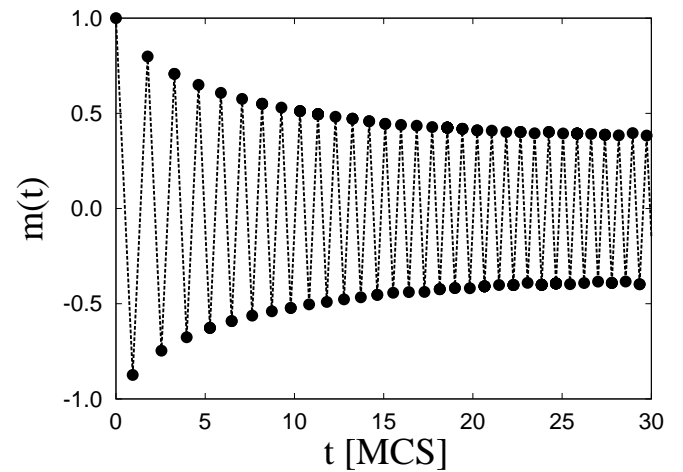}
\caption{
Magnetization profile for a typical sweep in the Ising model on a 
$64,000 \times 64,000$ square lattice using the Wolff algorithm 
at $T=2.269 J/k_{\rm. B}$.}
\label{fig1}
\end{figure}
\medskip
\par
{\it Formulation.}
In the present Letter, we concentrate on the 
two-dimensional Ising model on a square lattice. 
Numerical calculations are started from the all-up state using 
the Wolff algorithm,\cite{Wolff1} in which a single cluster is grown 
from a randomly chosen site up to a maximum size and is always 
flipped for the Ising model. The magnetization profile for a typical 
Monte Carlo sweep is shown in Fig.\ \ref{fig1}, where a 
$64,000 \times 64,000$-spin system is calculated at 
$T=2.269\ [J/k_{\rm B}] < T_{\rm c}
=2/\log (1+\hspace{-1.0mm}\sqrt{2})\ [J/k_{\rm B}] 
= 2.2691853\cdots [J/k_{\rm B}]$. 
In conventional Monte Carlo algorithms, all the spins are updated once 
in each Monte Carlo step (MCS), whereas only a single spin cluster 
is updated in each step in the Wolff algorithm. Then, the definition of 
a MCS in one step comparable to that in conventional algorithms is 
``(updated spin number) / (total spin number)". 

The ordered region is characterized by a spin cluster as large as the 
system size, and the bulk magnetization only originates from this cluster. 
Other spin clusters are much smaller and the magnetization after spatial 
averaging vanishes. Then, as shown in Fig.\ \ref{fig1}, the value 
of magnetization is almost reversed when the updated cluster is of 
the system size. On the other hand, the value is almost unchanged 
when the updated cluster is not so huge. The magnetization profile 
is completely governed by such a system-size cluster, and 
the absolute value of magnetization changes continuously. 
Therefore, magnetization profiles for different random-number 
sequences (RNSs) can be averaged in the following manner.
\begin{enumerate}
\item Take the absolute value $|m(t)|$ for each RNS.
\item Extract data obtained from system-size-cluster update.
\item Divide relaxation time with a small enough window $\Delta t$.
\item Average data for each time window $n \Delta t \le t < (n+1) \Delta t $ 
($n=0,1,2,\ldots$) to obtain $\langle |m(t)| \rangle$. 
\end{enumerate}

\begin{figure}[b]
\includegraphics[width=88mm]{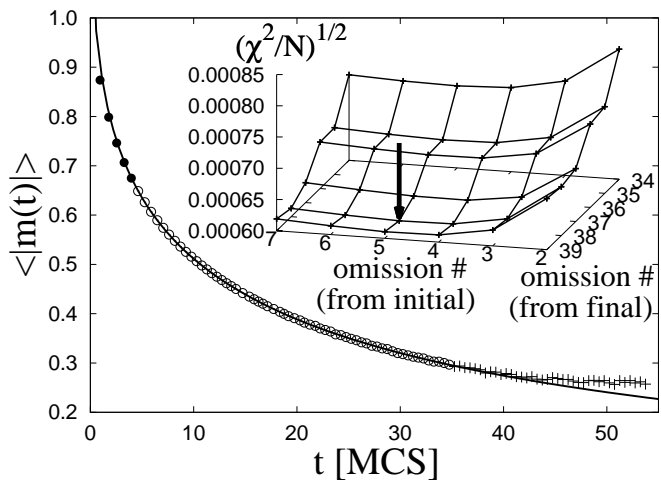}
\caption{
Fitting curve of average magnetization at $T=T_{\rm c}$ 
based on the stretched-exponential scaling form (\ref{se-SC}). 
Data points plotted by filled circles or crosses are not used 
for fitting. In the inset, the optimal number of omitted 
data points (from the two ends of the data sequence) 
is determined at the minimum residue per data 
$\chi^{2}/N$, as represented by the arrow.}
\label{fig2}
\end{figure}
\medskip
\par
{\it Simulation-time scaling for the largest system.}
We first analyze the average magnetization $\langle |m(t)| \rangle$ 
for the $64,000 \times 64,000$-spin system and $800$ 
RNS with $\Delta t=0.5$ MCS. We test various physical 
$2$-parameter scaling forms such as the power-law or 
exponential decay, and find that none of these forms can 
explain the present relaxation data very well. Then, it would be 
natural to consider $3$-parameter scaling forms, even though 
the use of fitting functions with many free parameters 
is dangerous; wide degrees of freedom may result in 
ambiguous fitting regardless of functional form. Taking these 
things into account, we test the stretched-exponential form 
\begin{equation}
\label{se-SC}
\langle |m(t)| \rangle \sim C \exp \left[ - ( t / \tau )^{\sigma} \right],\ 
0<\sigma <1,
\end{equation}
which is already established in local-update algorithms 
below $T_{\rm c}$,\cite{Huse-Fisher,TNM} where 
$\langle |m(t)| \rangle$ converges to the spontaneous 
magnetization $m_{\rm s}$. Equation (\ref{se-SC}) 
is a special case with $m_{\rm s}=0$. 

When we adopt a sufficiently long simulation time, 
$\langle |m(t)| \rangle$ obtained in the present framework 
is expected to diverge from the scaling form (\ref{se-SC}) 
and converge to the critical magnetization $m_{\rm c}(L)$, 
which vanishes as $m_{\rm c}(L) \sim L^{-\beta/\nu}$ for 
$L \to \infty$, with the critical exponents of spontaneous 
magnetization $\beta$ and correlation length $\nu$. 
Fitting the data with Eq.\ (\ref{se-SC}), we have 
\begin{equation}
\label{est}
\sigma=0.342 \pm 0.003,\ 
\tau=9.3 \pm 0.3,\ 
C=1.43 \pm 0.02.
\end{equation}
The fitting curve is displayed in Fig.\ \ref{fig2}, where the 
data points plotted by filled circles (initial ones) or crosses 
(longer-time ones) are not used for fitting. 
Since this calculation starts from $\langle |m(t=0)| \rangle=1$, 
which corresponds to $C=1$ in Eq.\ (\ref{se-SC}), several initial 
data points should be omitted to obtain a better fitting curve. 
As explained above, longer-time data points deviate from 
Eq.\ (\ref{se-SC}) and converge to $m_{\rm c}(L)$, and 
therefore should also be omitted. 
The number of omitted data points can be optimized 
systematically by minimizing the residue per data $\chi^{2}/N$, 
as shown in the inset of Fig.\ \ref{fig2}. Since systematic or 
statistical errors are dominant for smaller or larger number of 
omitted data points, respectively, the quantity $\chi^{2}/N$ 
is generally expected to have a minimum in between.

\begin{figure}[b]
\includegraphics[width=86mm]{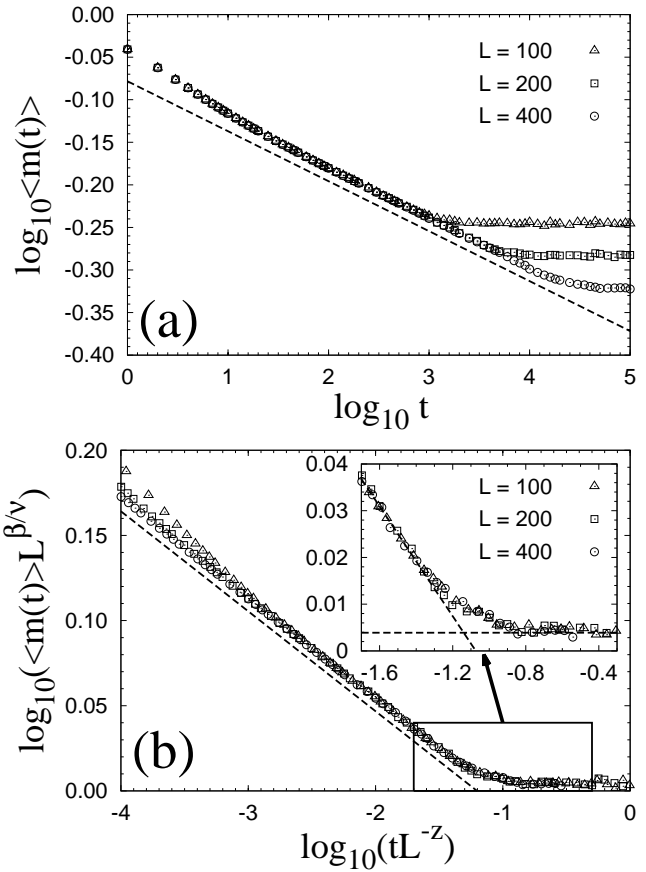}
\caption{
(a) Time dependence of the average magnetization based on 
the Metropolis algorithm for $100 \times 100$- (triangles), 
$200 \times 200$- (squares), and $400 \times 400$- (circles) 
spin systems in a log-log plot. Tangent of the broken line 
corresponds to $-\beta/(z \nu)$. (b) The scaling plot explained 
in the text. Data in the off-scaling region are enlarged in the inset, 
where the broken lines represent the power-law decay 
and equilibrium value and act as guides for the eyes.
}
\label{fig3}
\end{figure}
\medskip
\par
{\it New nonequilibrium-to-equilibrium dynamical scaling.}
The above results may be insufficient as evidence 
of the critical scaling form (\ref{se-SC}), because 
even the system with $64,000 \times 64,000$ spins 
may not be sufficiently large and the power-law region 
may appear for larger system sizes. 
Hence, we propose a new procedure to represent nonequilibrium 
and equilibrium regions in a panoramic manner and show that the 
relaxation behavior in Fig.\ \ref{fig2} is not a small-size effect. 

This procedure is first explained in the relaxation process in the 
Metropolis algorithm, a representative local-update algorithm. 
The time dependence of the average magnetization relaxed from 
the all-up state is displayed in Fig.\ \ref{fig3}(a) on a log-log scale 
for $100 \times 100$-, $200 \times 200$-, and $400 \times 400$-spin 
systems represented by triangles, squares, and circles, respectively. 
Each data point is obtained from 6,400 RNS, where the number of 
RNS remains unchanged while the system size increases, because 
the fluctuating nonequilibrium region expands simultaneously. 
The power-law behavior $\langle m(t) \rangle \sim t^{-\beta/(z \nu)}$ 
is clearly observed even in such a small system as for the 
standard NER analysis. Namely, the tangent of the broken 
line is $-\beta/(z \nu)$, with the exact exponent 
$\beta/\nu=1/8$ and the dynamical exponent 
$z \approx 2.13$ evaluated from these data.

In order to put all the data on a single curve, the size 
dependence of $\langle m(t) \rangle$ in equilibrium 
should be cancelled by multiplying $L^{\beta/\nu}$. 
Since this factor violates the size independence of 
data in the power-law region, the simulation time 
should be multiplied by $L^{-z}$ for compensation,  
i.e.~$(tL^{-z})^{-\beta/(z \nu)}=t^{-\beta/(z \nu)}L^{\beta/\nu}$. 
Then, we have a new scaling plot as shown in Fig.\ \ref{fig3}(b), 
where the tangent of the scaled data in the power-law region 
is the same as that in Fig.\ \ref{fig3}(a) as visualized by the 
broken line, and the scaled data in equilibrium fall on a 
horizontal line. It is interesting that the data between 
these two regions are also scaled on a single curve, 
and that such an intermediate region is quite narrow, 
as emphasized in the inset. 

\begin{figure}[b]
\includegraphics[width=86mm]{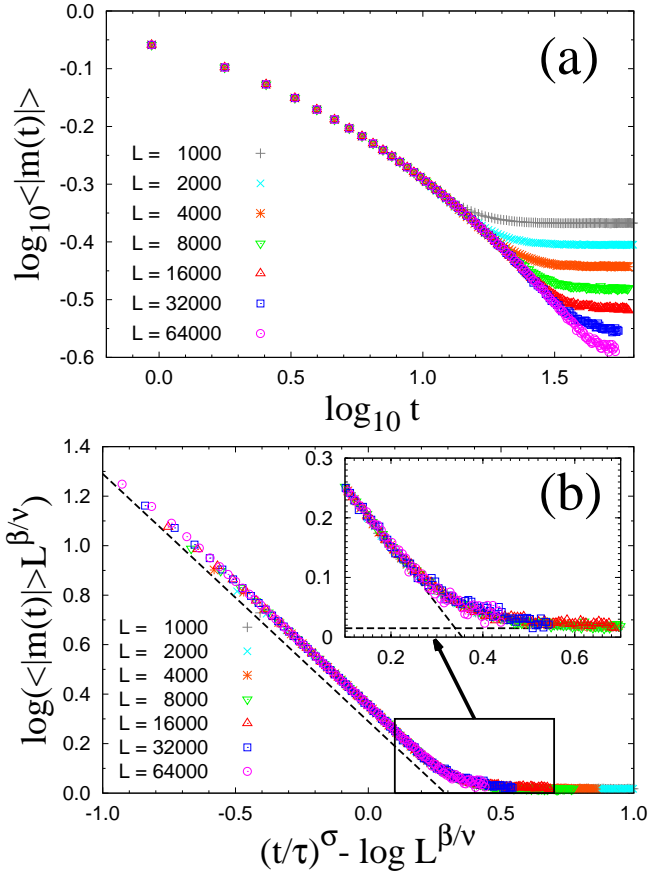}
\caption{
(Color online) 
(a) Time dependence of the average magnetization based on 
the Wolff algorithm for $1,000 \times 1,000$- (crosses), 
$2,000 \times 2,000$- (X-marks), $4,000 \times 4,000$- (stars), 
$8,000 \times 8,000$- (reverse-triangles), $16,000 \times 
16,000$- (triangles), $32,000 \times 32,000$- (squares), and
$64,000 \times 64,000$- (circles) spin systems on a log-log scale. 
(b) Scaling plot explained in the text. The tangent of the broken 
line is $-1$. Data in the off-scaling region are enlarged in the inset, 
where the broken lines represent the stretched-exponential decay 
and equilibrium value and act as guides for the eyes.
}
\label{fig4}
\end{figure}
Next, the average magnetization based on the Wolff algorithm 
for various system sizes is analyzed similarly. In Fig.\ \ref{fig4}(a), 
the data for $1,000 \times 1,000$- (409,600 RNS), $2,000 \times 
2,000$- (102,400 RNS), $4000 \times 4000$- (51,200 RNS), 
$8,000 \times 8,000$- (12,800 RNS), $16,000 \times 16,000$- 
(6,400 RNS), $32,000 \times 32,000$- (1,600 RNS), and 
$64,000 \times 64,000$- (800 RNS) spin systems are displayed 
on a log-log scale to enable comparison with Fig.\ \ref{fig3}(a). 
Apparently, the data do not exhibit a power-law decay, 
and the equilibration time is much shorter. 

The scaling plot corresponding to Fig.\ \ref{fig3}(b) can be derived as 
follows: (i) Multiply $L^{\beta/\nu}$ with $\langle |m(t)| \rangle$, because 
the equilibrium state does not depend on the relaxation process. 
(ii) In accordance with Eq.\ (\ref{se-SC}), $\exp[-(t/\tau)^{\sigma}]$ 
should be replaced by $\exp[-(t/\tau)^{\sigma}]L^{\beta/\nu}$ 
for compensation, or by 
$(t/\tau)^{\sigma} \rightarrow (t/\tau)^{\sigma}-\log L^{\beta/\nu}$, 
as displayed in Fig.\ \ref{fig4}(b) with 
the estimates given in Eq.\ (\ref{est}). 
Data for a wide range of system sizes are scaled well 
with the parameters obtained from the largest system. 
In this case, the off-scaling region is also quite narrow, 
as shown in the inset, which strongly suggests that no 
intermediate (i.e.\ power-law) region would exist between 
the stretched-exponential and equilibrium regions. 

\begin{figure}[b]
\includegraphics[width=88mm]{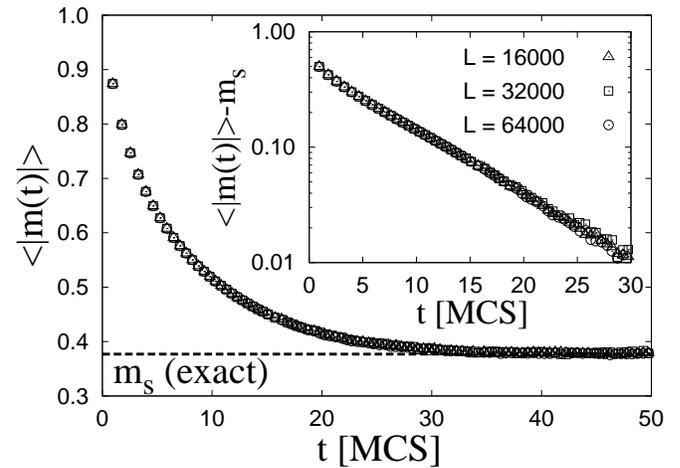}
\caption{
Time dependence of the average magnetization using the 
Wolff algorithm at $T=2.269 J/k_{\rm B}$, which is slightly below 
$T_{\rm c}$, for $16,000 \times 16,000$- (triangles), $32,000 
\times 32,000$- (squares), and $64,000 \times 64,000$- (circles) 
spin systems. Convergence to the exact spontaneous 
magnetization $m_{\rm s}$ is almost independent of 
system size, and the average magnetization subtracted 
by $m_{\rm s}$ decays exponentially, as shown in the 
inset (almost-linear semilog plot).
}
\label{fig6}
\end{figure}
\medskip
\par
{\it Relaxation behavior slightly below $T_{\rm c}$.}
Finally, results in the ordered phase are briefly reported. 
In simulations based on local-update algorithms, the 
stretched-exponential decay with $\sigma=1/2$ was 
confirmed\cite{ItoNER,Stauffer} in the Ising model on 
a square lattice. Then, similar calculations based on the 
Wolff algorithm are performed at $T=2.269 J/k_{\rm B}$, 
which is slightly below $T_{\rm c}$ and 
the same as that in Fig.\ \ref{fig1}. 
The average magnetization is plotted versus 
simulation time for $16,000 \times 16,000$- (800 RNS), 
$32,000 \times 32,000$- (100 RNS), and $64,000 \times 
64,000$- (100 RNS) spin systems in Fig.\ \ref{fig6}. 
These data rapidly converge to the exact value of 
the spontaneous magnetization $m_{\rm s}$, and 
the size dependence is negligible in this figure. 
In the inset, the average magnetization subtracted 
by $m_{\rm s}$ is plotted versus simulation time on 
a semilog scale. All the data fall on a straight line, 
that is, this quantity decays exponentially. 
\medskip
\par
{\it Discussion.}
In the stretched-exponential decay  in the Ising model below 
$T_{\rm c}$ based on local-update algorithms, the exponent 
is expressed as $\sigma=(d-1)/2$ with the spatial dimension 
$d$ in terms of the droplet picture,\cite{Huse-Fisher} and a 
similar analytic relation is also expected in the present case. 
According to our preliminary calculations for the Ising model 
on a cubic lattice at $T_{\rm c}$ based on the Wolff algorithm, 
the stretched-exponential decay with the exponent 
$\sigma \approx 0.45 \sim 0.5$ is observed. Our estimate for 
$d=2$ in Eq.\ (\ref{est}) and this value for $d=3$ might be 
consistent with $\sigma=(d-1)/(d+1)$,\cite{TNM} which was 
originally proposed for the region below $T_{c}$ but was not 
consistent with numerical studies.\cite{ItoNER,Stauffer} 
Further investigations are still necessary. 

The scaling plot in Figs.\ \ref{fig3}(b) and \ref{fig4}(b) 
reveals that each axis of the plot is scale-invariant. 
That is, Fig.\ \ref{fig3}(b) shows that the simulation time 
in local-update algorithms is scaled as $t \sim L^{z}$ 
at $T=T_{\rm c}$. Similarly, Fig.\ \ref{fig4}(b) indicates 
that the simulation time in cluster algorithms is scaled 
as $t \sim (\log L)^{1/\sigma}$ at $T=T_{\rm c}$. 

Studies of the dynamical critical exponent $z$ 
in cluster algorithms based on 
the equilibrium autocorrelation function of various 
quantities have a long history. Swendsen and Wang\cite{SW} 
and Wolff\cite{Wolff2} already made such calculations using 
their own algorithms and obtained $z$ much smaller than 
that obtained with local-update algorithms. 
That is, the autocorrelation function in a finite system of 
linear size $L$ decays exponentially with the correlation 
time $\tau(L)$, and the exponent $z$ is evaluated from 
the power-law scaling $\tau(L) \propto L^{z}$. 
Several studies followed along this 
line,\cite{Klein,Tamayo,Heermann,Baillie,Coddington} 
and it was found to be difficult to distinguish 
between the weak power-law and logarithmic 
size dependences of $\tau(L)$.\cite{Heermann,Baillie} 
Such a logarithmic scaling means $z=0$, namely, the 
breakdown of standard dynamical critical behaviors. 

Then, NER was introduced to investigate larger 
systems. G\"und\"uc {\it et al.}\cite{Gunduc} studied 
NER from the completely disordered state in the SW 
and Wolff algorithms, and reported $z=0$ in both cases. 
Du {\it et al.}\cite{Du} studied much larger systems, 
and found that NER from the completely ordered state 
gave $z=0$ for the SW and Wolff algorithms, whereas 
NER from the completely disordered state resulted in 
$z=0$ for the SW algorithm and $z \approx 1.19 $ for 
the Wolff algorithm. Recently, Liu {\it et al.}\cite{Liu} 
studied $z$ for the SW and Wolff algorithms using 
the Kibble-Zurek ansatz,\cite{Kibble,Zurek,KZrev} 
which is a kind of scaling theory based on the annealing 
process from a disordered state to $T_{\rm c}$. 
Critical exponents can be directly obtained from the size 
dependence of $\langle m^{2}(L) \rangle$, and they gave 
a small but finite $z$ comparable to that derived using 
equilibrium autocorrelation functions. 

In a word, the exponent $z$ in cluster algorithms is not yet 
established even today. Although our results in this Letter 
appear to be consistent with Ref.\ 5, we do not intend to claim 
$z=0$. Rather, we consider that the stretched-exponential 
decay from the perfectly ordered state is a specific property 
of cluster algorithms, and that this property would be useful 
when $z$ is still large even in cluster algorithms. 
Once the scaling form is established, NER analyses 
similar to the power-law case can be constructed. 

\medskip
\par
{\it Summary and future tasks.}
We have investigated early-time nonequilibrium 
relaxation behaviors based on the Wolff algorithm 
numerically in the two-dimensional Ising model on a 
square lattice, and found that they are totally different 
from those based on local-update algorithms. 
That is, the average magnetization decays 
in a stretched-exponential law at the critical 
temperature $T_{\rm c}$ and in an exponential 
law below $T_{\rm c}$ in cluster algorithms, 
whereas it decays in a power law at $T_{\rm c}$ and 
in a stretched-exponential law below $T_{\rm c}$ 
in local-update algorithms. 
We have proposed a new scaling procedure to draw nonequilibrium 
and equilibrium data for various system sizes on a single curve, 
and found that this scheme holds well for the power-law decay 
in the Metropolis algorithm and the stretched-exponential 
decay in the Wolff algorithm. Crossover regions between 
the scaling and equilibrium regions appear narrow in both 
cases, which suggests the absence of an intermediate region.

There are many future tasks within the present framework. 
First, a scheme comparable to the standard NER method should 
be established. A procedure for evaluating $T_{\rm c}$ should be 
obtained, and a formalism for evaluating critical exponents 
without equilibrium data is required. Second, the formalism 
should be generalized for the first-order or BKT phase transitions. 
Third, essential applications for the present framework should be 
specified. For example, relaxation in random spin systems is too 
slow to evaluate by the standard NER method based on 
local-update algorithms, but the present approach based 
on cluster algorithms might be effective.

Finally, extension to the quantum Monte Carlo (QMC) 
method is promising. Although 
modern QMC algorithms based on the continuous imaginary-time 
formalism\cite{Beard} have naturally been coupled with the cluster-update 
scheme,\cite{Evertz} NER analyses so far have been based on the world-line 
local update\cite{YN98} or on a special algorithm with cluster update only 
in imaginary time\cite{Nakamura} in order to retain the power-law decay. 
The present scheme enables NER analyses of QMC algorithms with cluster 
update, the code for which is faster and simpler than previous ones.

Investigations along these lines are now in progress.

\medskip
\par
{\it Acknowledgement.}
Y.~N.\ would like to thank Y.~Tomita for stimulating 
discussion. He also thanks M.~Kohno, S.~Todo and Y.~Ozeki for 
helpful comments. The random-number generator 
MT19937\cite{MT} was used for numerical calculations.

\end{document}